\documentclass{svmult}

\usepackage{makeidx}         % allows index generation
\usepackage{graphicx}        % standard LaTeX graphics tool
\usepackage{multicol}        % used for the two-column index
\usepackage[bottom]{footmisc}% places footnotes at page bottom
\makeindex             % used for the subject index
\begin{document}

\title*{Time series of stock price and of two fractal overlap:
Anticipating market crashes?}
\titlerunning{Two fractal overlap: Anticipating market crashes?}
\author{Bikas K. Chakrabarti\inst{1}, Arnab Chatterjee\inst{2} \and
Pratip Bhattacharyya\inst{3}}
\institute{Theoretical Condensed Matter Physics Division and Centre for Applied Mathematics and Computational Science, Saha Institute of Nuclear Physics, Block-AF, Sector-I Bidhannagar, Kolkata-700064, India.
\texttt{bikas@cmp.saha.ernet.in}
\and \texttt{arnab@cmp.saha.ernet.in}
\and \texttt{pratip@cmp.saha.ernet.in}}
\maketitle
%%%%%%%%%%%%%%%%%%%%%%%%%%%%%%%%%%%%%%%%%%%%%%%%%%%%%%%%%%%%%%%%%%%%%%%%%%%
We find prominent similarities in the features of the time series 
for the overlap of two Cantor sets when one set moves with uniform relative
velocity over the other and time series of stock prices. 
An anticipation method for some of the crashes have been proposed here,
based on these observations.
%%%%%%%%%%%%%%%%%%%%%%%%%%%%%%%%%%%%%%%%%%%%%%%%%%%%%%%%%%%%%%%%%%%%%%%%%%%

%%%%%%%%%%%%%%%%%%%%%%%%%%%%%%%%%%%%%%%%%%%%%%%%%%%%%%%%%%%%%%%%%%%%%%%%%%%
\section{Introduction}
\label{intro}
%%%%%%%%%%%%%%%%%%%%%%%%%%%%%%%%%%%%%%%%%%%%%%%%%%%%%%%%%%%%%%%%%%%%%%%%%%%

Capturing dynamical patterns of stock prices are
major challenges both epistemologically as well as financially
\cite{book}. The statistical properties of their (time) variations
or fluctuations \cite{book} are now well studied and
characterized (with established fractal properties), but are not very
useful for studying and anticipating their dynamics
in the market. Noting that a single fractal gives essentially a time
averaged picture, a minimal two-fractal overlap time series model was
introduced \cite{Chakrabarti:1999,Pradhan:2003,Pradhan:2004}.

%%%%%%%%%%%%%%%%%%%%%%%%%%%%%%%%%%%%%%%%%%%%%%%%%%%%%%%%%%%%%%%%%%%%
\section{The model}
\label{model}
%%%%%%%%%%%%%%%%%%%%%%%%%%%%%%%%%%%%%%%%%%%%%%%%%%%%%%%%%%%%%%%%%%%%

We consider first the time series $O(t)$ of the overlap sets of two identical
fractals \cite{Pradhan:2004,Bhattacharyya:2005}, as one slides over the 
other with uniform velocity. Let us consider two regular cantor sets
at finite generation $n$. As one set slides over the other, the overlap set
changes. The total overlap $O(t)$ at any instant
$t$ changes with time
(see Fig. 1(a)). In Fig. 1(b) we show the behavior of the cumulative overlap
\cite{Pradhan:2004} $Q^o(t) = \int_0^t O(\tilde{t}) d\tilde{t}$.
This curve, for sets with generation $n=4$,
is approximately a straight line \cite{Pradhan:2004} with slope $(16/5)^4$.
In general, this curve approaches a strict straight line in the limit
$a \rightarrow \infty$, asymptotically, where the overlap set comes from the
Cantor sets formed of $a-1$ blocks, taking away the central block,
giving dimension of the Cantor sets equal to $\mathrm{ln}(a-1)/\mathrm{ln}a$.
The cumulative curve is then almost a straight line and has then a slope
$\left[(a-1)^2/a\right]^n$ for sets of generation $n$.
If one defines a `crash' occurring at time $t_i$ when
$O(t_i)-O(t_{i+1}) \ge \Delta$ (a preassigned large value) and one
redefines the zero of the scale at each $t_i$,
then the behavior of the cumulative overlap
$Q^o_i(t) = \int_{t_{i-1}}^t O(\tilde t) d \tilde{t},\; \tilde{t} \le t_i$,
has got the peak value `quantization' as shown in Fig. 1(c). The reason
is obvious. This justifies the simple thumb
rule: one can simply count the cumulative
$Q^o_i(t)$ of the overlaps since the last `crash' or `shock' at $t_{i-1}$
and if the value exceeds the minimum value ($q_o$), one can safely extrapolate
linearly and expect growth upto $\alpha q_o$ here and face a `crash' or overlap
greater than $\Delta$ ($=150$ in Fig. 1). If nothing happens there,
one can again wait upto a time until which the cumulative grows upto 
$\alpha^{2}q_o$ and feel a `crash' and so on 
($\alpha=5$ in the set considered in Fig. 1).

%%%%FIGURE%%%%%%%%%%%%%%%%%%%%%%%%%%%%%%%%%%%%%%%%%%%%%%%%%%%%%%%%%%
\begin{figure}
\centering
\resizebox*{9.5cm}{!}{\rotatebox{0}{\includegraphics{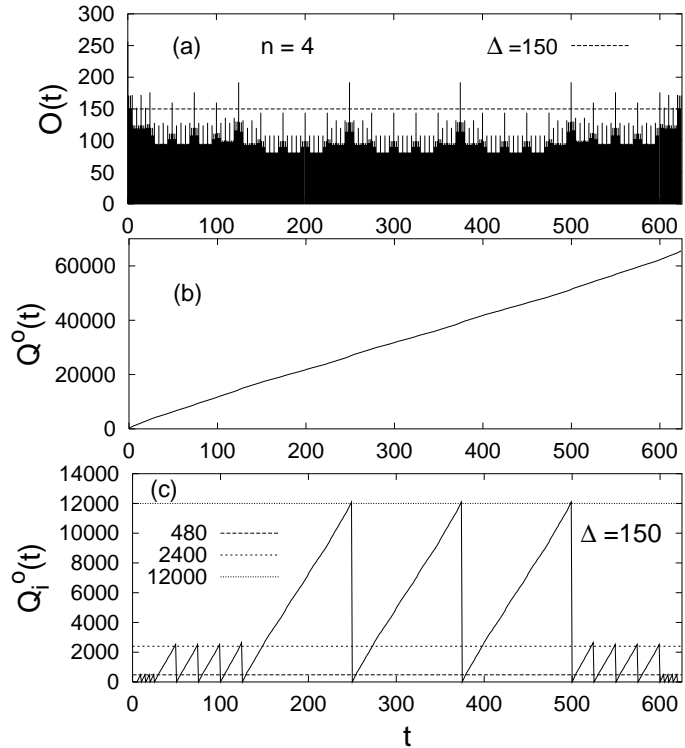}}}
\caption{(a) The time series data of overlap size $O(t)$ for a regular Cantor
set of dimension $\rm{ln}4/\rm{ln}5$ at generation $n=4$.
(b) Cumulative overlap $Q^o(t)$ and
(c) the variation of the cumulative overlap $Q^o_i(t)$ for the same series,
where $Q$ is reset to zero after any big event of size
greater than $\Delta=150$. }
\label{fig:1} 
\end{figure}
%%%%FIGURE%%%%%%%%%%%%%%%%%%%%%%%%%%%%%%%%%%%%%%%%%%%%%%%%%%%%%%%%%%

We now consider some typical stock price time-series data, available in the 
internet \cite{NYSE}. In Fig. 2(a), we show that the daily stock price $S(t)$
variations for about $10$ years (daily closing price of the `industrial index')
from January 1966 to December 1979 (3505 trading days). The
cumulative $Q^s(t) = \int_0^t S(t) dt$ has again a straight line
variation with time $t$ (Fig. 2(b)).
We then define the major shock by identifying those variations when
$\delta S(t)$ of the prices in successive days exceeded a preassigned
value $\Delta$ (Fig. 2(c)).
The variation of $Q_i^s(t) = \int_{t_{i-1}}^{t_i} S(\tilde{t}) d\tilde{t}$
where $t_i$ are the times when $\delta S(t_i) \le -1$ show similar
geometric series like peak values (see Fig. 2(d)).

%%%%FIGURE%%%%%%%%%%%%%%%%%%%%%%%%%%%%%%%%%%%%%%%%%%%%%%%%%%%%%%%%%%
\begin{figure}
\centering
\resizebox*{9.5cm}{!}{\rotatebox{0}{\includegraphics{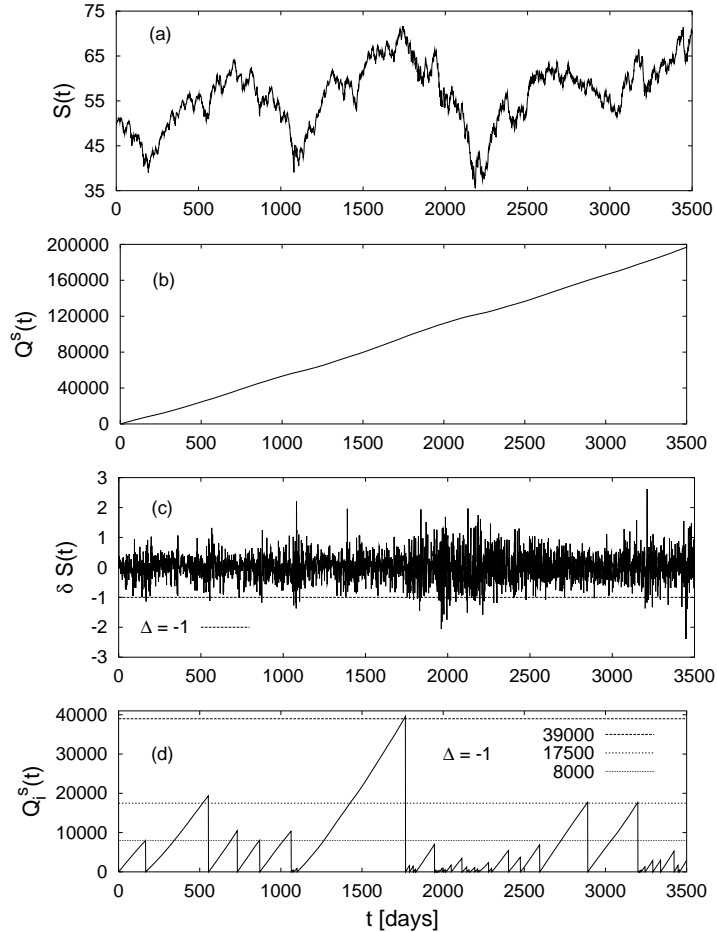}}}
\caption{Data from New York Stock Exchange from January 1966 to December 1979:
industrial index \cite{NYSE}:
(a) Daily closing index $S(t)$ (b) integrated
$Q^s(t)$,
(c) daily changes $\delta S(t)$ of the index $S(t)$ defined as
$\delta S(t) = S(t+1) - S(t)$, and (d) behavior of $Q_i^s(t)$
where $\delta S(t_i) > \Delta$. Here, $\Delta=-1.0$ as shown in (c) by
the dotted line.  }
\label{fig:2} 
\end{figure}
%%%%FIGURE%%%%%%%%%%%%%%%%%%%%%%%%%%%%%%%%%%%%%%%%%%%%%%%%%%%%%%%%%%

A simple `anticipation strategy' for some of the crashes may be as follows:
If the cumulative $Q_i^s(t)$ since the last crash has grown beyond
$q_0 \simeq 8000$ here, wait until it grows (linearly with time) until about
$17,500$ ($\simeq 2.2q_0$) and expect a crash there. If nothing happens,
then wait until $Q_i^s(t)$ grows (again linearly with time) to a value of the 
order of $39,000$ ($\simeq (2.2)^2 q_0$) and expect a crash, and so on.

%%%%%%%%%%%%%%%%%%%%%%%%%%%%%%%%%%%%%%%%%%%%%%%%%%%%%%%%%%%%%%%%%%%%
\section{Summary}
%%%%%%%%%%%%%%%%%%%%%%%%%%%%%%%%%%%%%%%%%%%%%%%%%%%%%%%%%%%%%%%%%%%%
The features of the time series for the overlap of two Cantor sets when one 
set moves with uniform relative velocity over the other looks somewhat similar
to the time series of stock prices. 
We analyze both and explore the possibilities of anticipating a large
(change in Cantor set) overlap or a large change in stock price.
An anticipation method for some of the crashes has been proposed here,
based on these observations.

%%%%%%%%%%%%%%%%%%%%%%%%%%%%%%%%%%%%%%%%%%%%%%%%%%%%%%%%%%%%%%%%%%%%%%%%%%%

%%%%%%%%%%%%%%%%%%%%%%%%%%%%%%%%%%%%%%%%%%%%%%%%%%%%%%%%%%%%%%%%%%%%%%%%%%%

\printindex
\end{document}